\documentstyle[preprint,aps]{revtex}
\def\be{\begin{equation}}
\def\ee{\end{equation}}
\begin{document}
\draft
\title{
Current-voltage characteristics of a tunnel junction with resonant centers
	 }
\author{T. Ivanov\cite{byline}$^{(a)}$ and V. Valtchinov$^{(b)}$}
\address{
$^{(a)}$ International Centre for Theoretical Physics, P.O.Box 586, I-34100,
Trieste, Italy \\
$^{(b)}$ Department of Physics,
Norhteastern University, Boston, MA 02115.
	  }
\maketitle
\begin{abstract}
We calculate the I-V characteristics of a tunnel junction containing
impurities in the barrier. We consider the indirect resonant tunneling
involving the impurities. The Coulomb repulsion energy $E_c$ between two
electrons with opposite spins simultaneously residing on the impurity is
introduced
by an Anderson Hamiltonian. At low temperatures $T \ll E_c$ the I-V
characteristic is linear in $V$ both for $V<E_c$ and for $V>E_c$ and changes
slope at $V=E_c$. This
behavior reflects the energy spectrum of the impurity electrons - the
finite value of the charging energy $E_c$. At $T \sim E_c$ the junction
reveals an ohmic-like behavior as a  result of the smearing out of the
charging effects by the thermal fluctuations.
\end{abstract}
In this paper, we study the transport through a tunnel junction with a
thin, amorphous insulating layer separating the metal leads. At low
temperatures two tunneling channels contribute to the conductance. The
first one is the direct tunneling between the leads. The second one is a
resonant tunneling through impurity states present in the barrier. This
mechanism is dominant for not too small thickness of the layer \cite{1,2}.
For higher temperatures and larger applied voltages the inelastic
tunneling involving pairs of
localized states with phonon emission or absorption becomes important
\cite{3,4,5}. We restrain our study to the case of resonant tunneling and
calculate the current-voltage characteristics of such junctions. The
current flows through the junction by tunneling of electrons from the
emitter electrode to the impurity and then to the collector. We calculate
the tunneling current through the impurity by taking into
account the Coulomb repulsion energy between two electrons
with opposite spins simultaneously
residing on the localized state. In systems such as {\it a-Si} tunnel
junctions this energy is rather large ($0.1 - 0.2 \ eV$) \cite{5,6}.
Experimentally measured current for a given {\it dc} voltage bias is
obtained by averaging the currents through the individual impurities over
their positions and energy distribution.

The Hamiltonian of the system under consideration consists of three terms:
\begin{equation}
H_{tot} = H_{leads} + H_{imp}+H_{tun}.
\end{equation}
The Hamiltonian describing the leads is written as:
\begin{equation}
H_{leads} = \sum\limits_{k\sigma}\epsilon_{k}^{L}a^{\dag}_{k\sigma}
a_{k\sigma}+
\sum\limits_{p\sigma}\epsilon_{p}^{R}b^{\dag}_{p\sigma}b_{p\sigma}.
\end{equation}
The impurity Hamiltonian is:
\begin{equation}
H_{imp}=\epsilon_{c} \sum _{\sigma} c_{\sigma}^{\dag}c_{\sigma}+
E_{c}n_{\uparrow}n_{\downarrow},
\end{equation}
and the tunneling term is:
\begin{equation}
H_{tun}=\sum\limits_{k\sigma}(T_{Lk}c^{\dag}_{\sigma}a_{k\sigma}+H.c.)+
      \sum\limits_{p\sigma}(T_{Rp}b^{\dag}_{p\sigma}c_{\sigma} + H.c.).
\label{tunnel}
\end{equation}
The Hamiltonian is expressed in terms of annihilation(creation) operators
$a_{k\sigma}$ for the emitter, $b_{p\sigma}$ for the collector and
$c_{\sigma}$ for the impurity with $k(p)$ the corresponding quasimomenta
and $\sigma = \uparrow, \downarrow$ is the spin index. $\epsilon^{L}_{k}$
and $\epsilon^{R}_{p}$ are the single-particle
energies in the emitter and the collector leads, respectively. $E_{c}$ is
the Coulomb repulsion between electrons with opposite spins. The
single-electron
energies are measured from the corresponding Fermi levels
$\mu_{L}$ and $\mu_{R}$ in the emitter and collector, and the $dc$
bias is $\mu_{L}-\mu_{R}=eV$.
The resonant level energy is $\epsilon_c=\epsilon^0_c+\alpha eV$, where
$\epsilon^0_c$ is the bare resonant level energy and $\alpha$ measures the
portion of the voltage drop on the localized center.
The particle-number operator is
$n_{\sigma}=c^{\dag}_{\sigma}c_{\sigma}$.
In Eqn.~(\ref{tunnel}) for the tunneling Hamiltonian, $T_{Lk}$ and $T_{Rp}$ are
the tunneling
matrix elements. Note that they depend on the impurity position in the
junction. We assume that the impurities are randomly distributed in the
junction barrier and their resonant levels are uniformly distributed in
energy.

First we calculate the impurity's Green's function. To correctly account
for the non-equilibrium nature of the system we apply the Keldysh
technique \cite{7}. In this technique one introduces the
retarded \ (advanced) and distribution Green's functions. We further assume
that the relaxation processes in the leads are much faster than in the
impurity. Thus, we can consider the leads as equilibrium systems and the
corresponding Green's functions are given by the usual expressions for a
non-interacting equilibrium electron systems \cite{7}. Here we give only
the final expressions for the impurity's Green's functions (for details
on the derivation see \cite{8}). The retarded Green's function is
obtained in the form
\FL
\ \\
\begin{equation}
G_{r} = {\omega - \tilde {\epsilon}_{c\sigma} -
\Sigma_0 - \Sigma_1 \over {(\omega - \epsilon_{c} - \Sigma_0)(\omega -
\epsilon_c - \Sigma_0 - \Sigma_1) - E_c (\omega - \epsilon_c - \Sigma_0 -
\Sigma_2) }},
\label{general}
\end{equation}
\ \\
where the analytical continuation $\omega \longrightarrow \omega +i0^+$
is used. The self-energy parts in Eqn.~(\ref{general}) are given by
%
\begin{equation}
\Sigma_{0}(\omega)=\sum_{k,i=L,R} \frac{|T_{ik}|^{2}} {\omega-\epsilon^{i}_{k}
},
\end{equation}
\ \\
%
\begin{equation}
\Sigma_{1}(\omega)=\sum_{k,i=L,R}|T_{ik}|^{2}
\left [
\frac{1} {\omega-\epsilon_{c-\sigma}-\epsilon_{c\sigma}+\epsilon^{i}_{k} } +
\frac{1} {\omega+\epsilon_{c-\sigma}-\epsilon_{c\sigma}-\epsilon^{i}_{k} }
     \right ],
\end{equation}
\ \\
%
\begin{equation}
\Sigma_{2}(\omega)=\sum_{k,i=L,R}|T_{ik}|^{2}f(\epsilon^{i}_{k})
\left [
\frac{1} {\omega+\epsilon_{c-\sigma}-\epsilon_{c\sigma}-\epsilon^{i}_{k} } +
\frac{1} {\omega-\epsilon_{c-\sigma}-\epsilon_{c\sigma}+\epsilon^{i}_{k} }
     \right ],
\label{se2}
\end{equation}
\ \\
where $f(\epsilon^{i}_{k})$  is the Fermi-Dirac distribution function for
the electrons in the leads. We have used the following
notations: $\epsilon_{c\sigma}=\epsilon_c+E_c <n_{-\sigma}>$ and
$\tilde{\epsilon}_{c\sigma}=\epsilon_c+E_c (1-<n_{-\sigma}>)$. This
Green's functions describes two levels for the impurity electrons - a
lower, resonant level with energy $\epsilon_c$ and the upper level with
energy $\epsilon_c + E_c$.

The distribution Green's function is calculated under the assumption that
the transient processes after the switching-on the {\it dc} bias have
decayed and the result is
\begin{eqnarray}
G_{<}(\omega)= &-& \frac{ \sum \limits_{k}
|T_{Lk}|^{2}A_{<}
(k, \omega) + \sum \limits_{p} |T_{Rp}|^{2}B_{<}(p, \omega)}  {\sum\limits_{k}
|T_{Lk}|^{2}(A_{r}(k, \omega)-A_{a}(k, \omega)) +\sum\limits_{p}|T_{Rp}|^{2}
(B_{r}(p, \omega) - B_{a}(p, \omega) ) }\times \nonumber \\
&& \left( G_r(\omega) - G_a(\omega) \right),
\label{distr}
\end{eqnarray}
where $A_{r,a,<}(B_{r,a,<})$ are the retarded, advanced, and distribution
Green's functions in the left (right) lead.

The average number of impurity electrons is calculated by solving the
following equation
\be
n=-\int \frac {d\omega}{2\pi} Im G_{<}(\omega)
\label{nn}
\ee
(we assume $<n_{\sigma}> = <n_{-\sigma}> = n$).

It should be mentioned
(\cite{9} and references therein) that the Green's function formalism
presented
here is valid only for temperatures higher than the characteristic temperature
for this problem -- the Kondo temperature $T_K$.

Next we calculate the current through the impurity. In a steady-state it
is given by
\be
I(\epsilon_c) = i e \int \frac {d\omega} {2\pi} \frac {2 \gamma_L(\omega)
\gamma_R(\omega)}
{ \gamma_L(\omega) + \gamma_R(\omega)} \left( f_L(\omega) - f_R(\omega)
\right) \left ( G_r(\omega) - G_a(\omega) \right ),
\ee
where
\begin{equation}
\gamma_L(\omega)= \pi \sum\limits_{k}|T_{Lk}|^{2}
\delta(\omega-\epsilon^{L}_{k}), \hspace{0.3cm}
\gamma_R(\omega)= \pi \sum\limits_{p} |T_{Rp}|^{2} \delta(\omega -
\epsilon^{R}_{p}) \label{gami}
\end{equation}
are the elastic widths of the impurity level due to tunneling
through the left (right) barrier, respectively.
We take broad, flat density of states for the leads' electrons and
in the following $\gamma_L(\omega)$ and $\gamma_R(\omega)$ are
assumed to be constants, $\gamma_L$ and $\gamma_R$, independent of energy.

Now we compare our treatment with the work of Glazman and Matveev \cite{6}.
Their results can be obtained from ours by approximating the well electrons
Green's function (Eqn. ~(\ref{general})) with the corresponding expression
for an isolated one-center Hubbard model \cite{hub} {\it i.e. } one has to
 set all the self-energies $\Sigma_i (\omega),\ i=0, 2$ (Eqns. (6-8)) to
zero. This approach neglects the
virtual tunnel processes and it is justifiable when the temperature $T$
is much larger than the elastic level width $\gamma = \gamma_L +
\gamma_R$. With this approximation  the integral equation for the average
number of well electrons (Eqn. 10) reduces to an algebraic one whose
solution coincides with Eqn. 6 in Ref. 6.
Similarly, formula (11) for the current gives the simple analytical expression
Eqn. 8 in Ref. 6.

To obtain the current through the junction we must integrate over the
position of the impurity and to average over the energies of the
impurity levels. If we denote the position of the impurity, relative to the
left lead, by $x$ then $\gamma_L$ and $\gamma_R$ depend exponentially on $x$
\be
\gamma_L = \gamma_0 \exp \left ( - \frac {2x} { \xi} \right ),
\hspace{0.5cm} \gamma_R = \gamma_0 \exp \left ( - \frac {2(d-x)} {\xi}
\right ),
\label{zax}
\ee
where $d$ is the width of the tunnel junction and $\xi$ is the impurity
localization length. In a first approximation the coefficient $\gamma_0$
can be considered independent on $x$. One can show that most significant
contribution to the current through the junction give the impurities which
are close to its middle - $x = d/2$. Therefore, we simplify the
calculation by taking $x = d/2$ in Eqn.~(\ref{zax}) and discarding the
impurities which are far from this optimal position. With this
approximation the current through the junction is given by
\be
I = S d \int d\epsilon_c g(\epsilon_c) I(\epsilon_c)
\label{tok}
\ee
where $S$ is the area of the junction and $g(\epsilon_c)$ is the density
of impurity states. In further calculations we take $g(\epsilon_c)$ to be
a constant. In Eqn. ~(\ref{tok}) the integration is over all impurities
which give non-zero contribution to the current.

In Fig. 1 we show the current through the impurities as a function of the
energy of the impurity level for: a) $V=2.2 E_c$; b) $V=0.6E_c$  and
for $T=0.08
E_c$ $(\gamma_0 = 0.02 E_c)$. We have set $\mu_R = 0$ since the results
are independent on the value of the right chemical potential. The current is
almost zero when both
impurity levels - the resonant and the upper level, are well below the
right chemical potential or well above the left chemical potential. In
the former case $n=1$ - there are 2 electrons on the impurity and in the
latter $n \approx 0$. For $V>E_c$ the current is maximal when both
impurity levels are between the left and right chemical potentials - there
is approximately 1 electron on the impurity. The left
"shoulder" in $I(\epsilon_c)$ dependence for $V=2.2 E_c$ (Fig. 1 a)
corresponds to impurities for which $n \approx 1/3$ and the right
"shoulder" is due to impurities with $n \approx 2/3$. For $V<E_c$ there
are two peaks in the $I(\epsilon_c)$ dependence (Fig. 1 b). The lower
comes from tunneling through the upper level for impurities deep below
$\mu_R$ (for them only the upper level is above $\mu_R$). The upper peak
is due to impurities whose resonant level is above $\mu_R$ and the upper
level is above $\mu_L$. In this case the tunneling current flows
predominantly through the resonant level. In Fig. 1 c we show
$I(\epsilon_c)$ dependence for $T=0.9 E_c$ and $V=0.6 E_c$. It is
evident that for temperatures comparable with $E_c$ all charging effects
are smeared out by the thermal fluctuations.

In Fig. 2 we present the I-V characteristic of the tunnel junction,
calculated for: a) $T=0.002E_c$ and b) $T=0.9E_c$. In the former case
the current is linear in $V$ both for low voltage $V<E_c$ and for
$V>E_c$ and changes slope at $V=E_c$ - the slope diminishes for $V>E_c$.
This behaviour can qualitatively be understood by the following
considerations. With increasing $V$ more impurities are involved in the
transport through the junction. As $V$ increases to $E_c$ the current
through the junction increases significantly since it flows through
impurities which give maximal current (Fig. 1). Increasing $V$ above
$E_c$ includes impurities with resonant levels close to $\mu_L$ - for
them the current is rapidly decreasing function of $\epsilon_c$ (Fig.
1). We must point out that for temperatures $T<0.5 E_c$ the I-V
characteristic is practically independent on the temperature which again
is a signature of the charging effects.

The I-V characteristic at $T=0.9 E_c$ (Fig. 2b) shows ohmic behavior of
the tunnel junction. The current is linear in $V$ for every voltage -
the charging effects of the resonant tunneling are fully masked by the
thermal fluctuations.

In this work we have approximated the tunneling matrix elements to be
independent on energy (equivalently, on the applied {\it dc} voltage). One
may speculate that taking $T_{L(R)}$ to be increasing functions of $V$
and $\epsilon_c$ (softening of the barrier with increasing energy) will
give the experimentally observed I-V characteristics \cite{3} - for $V$
exceeding $E_c$ the slope of the $I(V)$ dependence increases.

In summary, in the present paper the phenomenon of resonant tunneling through
a tunnel junction containing impurity states in the barrier region
has been studied with the Coulomb repulsion between
electrons on the impurity taken into account. The I-V characteristic at
temperature $T \ll E_c$ is linear in V both for $V<E_c$ and for $V>E_c$
and its slope diminishes at $V=E_c$. The tunnel junction has ohmic-like
I-V characteristic for temperatures of the order of the charging energy
when the charging effect is smeared out by the thermal fluctuations.

\acknowledgements

The authors are thankful to Dr. A. Groshev for the stimulating
discussions. The work of T. I. at the International Centre for
Theoretical Physics was supported by a fellowship granted by the
Commission of the European Communities.
\begin{figure}
\caption{
 Current (in arb. units) through the impurity centers as a
	function of their energy $\epsilon_c$  for
 $T=0.08E_c$: a) $V=2.2 E_c$ (the dotted line), b) $V=0.6 E_c$ (the
solid line), and c) for $T=0.9 E_c$ and $V=0.6 E_c$ (the dashed line).
   }
\end{figure}
\begin{figure}
\caption{
 I-V characteristics of the tunnel junction for: a) $T=0.002 E_c$ (the
solid line), and b) $T=0.9 E_c$ (the dashed line). The current is
normalized to $I(E_c)$.
  }
\end{figure}

\end{document}